\documentclass[prl,%superscriptaddress,
twocolumn,letterpaper,%
nofootinbib,showpacs,preprintnumbers,amsmath,amssymb]{revtex4}
\usepackage{bm,curves}

\begin{document}

\title{
Gluon polarisation in the proton
}
\author{Steven D. Bass}
\affiliation{
Institute for Theoretical Physics, %\\
Universit\"at Innsbruck,
Technikerstrasse 25, Innsbruck, A 6020 Austria}
\author{Andrew Casey}
\affiliation{
CSSM, School of Chemistry and Physics, %\\
University of Adelaide, Adelaide SA 5005, Australia}
\author{Anthony W. Thomas}
\affiliation{
CSSM, School of Chemistry and Physics, %\\
University of Adelaide, Adelaide SA 5005, Australia}

\begin{abstract}
\noindent
We combine heavy-quark renormalisation group arguments with our 
understanding of the nucleon's wavefunction to deduce a bound on 
the gluon polarisation $\Delta g$ in the proton.
The bound is consistent with the values extracted from spin
experiments at COMPASS and RHIC.
\end{abstract}
\maketitle

\vfill\eject

Polarised deep inelastic scattering experiments have revealed a small 
value for the nucleon's flavour-singlet axial-charge,
$g_A^{(0)}|_{\rm pDIS} \sim 0.3$,
suggesting that 
the quarks' intrinsic spin contributes little of the proton's spin.
The challenge to understand the spin structure of the 
proton~\cite{bassrmp,bassbook,mpla,Thomas:2008bd,Thomas:2008ga,reya}
has inspired a vast programme of theoretical activity and new experiments.
Why is the quark spin content $g_A^{(0)}|_{\rm pDIS}$ so small ?
How is the spin $\frac{1}{2}$ of the proton built up from 
the spin and orbital angular momentum of the quarks and gluons inside ?

A major topic of investigation has been the role of polarised glue in
the nucleon, both in terms of its contribution to the nucleon's spin and 
possible supression of the nucleon's singlet axial-charge through 
the QCD axial anomaly.
Key experiments to measure gluon polarisation are COMPASS at CERN and
PHENIX and STAR at RHIC.
In this paper we investigate gluon polarisation 
via the charm-quark axial-charge, 
matching the results of 
heavy-quark renormalisation group 
with what we know of the proton's wavefunction.
We 
suggest a bound $|\Delta g (m_c^2)| \lesssim 0.3$, 
which is consistent with the results of the present experiments.

We start by recalling the $g_1$ spin sum-rules, which  
are derived from the dispersion relation for polarised
photon-nucleon scattering and, for deep inelastic scattering,
the light-cone operator product expansion.
At leading twist
the first moment of  the $g_1$ spin structure function measures a 
linear combination of the nucleon's scale-invariant axial-charges
$g_A^{(3)}$, $g_A^{(8)}$ and $g_A^{(0)}|_{\rm inv}$ 
plus a possible
subtraction constant $\beta_{\infty}$
in the dispersion relation \cite{bassrmp}:
\begin{eqnarray}
\int_0^1 dx \ g_1 (x,Q^2) &=& 
\Biggl( \frac{1}{12} g_A^{(3)} + \frac{1}{36} g_A^{(8)} \Biggr)
c_{\rm NS} 
(\alpha_s (Q^2))
\nonumber \\
& &
+ \frac{1}{9} g_A^{(0)}|_{\rm inv}
c_{\rm S} 
(\alpha_s (Q^2))
 + \ \beta_{\infty}.
\label{eqc50}
\end{eqnarray}
Here $c_{\rm NS}$ and $c_{\rm S}$ are the non-singlet and singlet
Wilson coefficients. 
In terms of the flavour dependent axial-charges
$2M s_{\mu} \Delta q =
\langle p,s |
{\overline q} \gamma_{\mu} \gamma_5 q
| p,s \rangle
$
the isovector, octet and singlet axial charges are
$g_A^{(3)} = \Delta u - \Delta d$,
$g_A^{(8)} = \Delta u + \Delta d - 2 \Delta s$
and
$
g_A^{(0)}|_{\rm inv}/E(\alpha_s) 
\equiv 
g_A^{(0)} (Q^2)
= \Delta u + \Delta d + \Delta s$.
Here
$
E(\alpha_s) = \exp \int^{\alpha_s}_0 \! 
d{\tilde \alpha_s}\,
\gamma({\tilde \alpha_s})/\beta({\tilde \alpha_s})
$
is a renormalisation group factor which 
corrects for the (two loop) non-zero anomalous dimension 
$\gamma(\alpha_s)$
of the singlet axial-vector current~\cite{kodaira},  
$
J_{\mu5} = 
\bar{u}\gamma_\mu\gamma_5u
                  + \bar{d}\gamma_\mu\gamma_5d
                  + \bar{s}\gamma_\mu\gamma_5s %\right)_{GI}
$ ,
which is close to one and which goes to one in the limit 
$Q^2 \rightarrow \infty$;
$\beta (\alpha_s) 
= - \biggl(11 - \frac{2}{3} f\biggr) (\alpha_s^2 / 2 \pi) + ...$
is the QCD beta function
and
$\gamma(\alpha_s) = f (\alpha_s / \pi)^2 + ...$
where $f$ is the number of active flavours. 
The singlet axial charge, $g_A^{(0)}|_{\rm inv}$,
is independent of the renormalisation scale $\mu$
and corresponds
to 
$g_A^{(0)}(Q^2)$ evaluated in the limit $Q^2 \rightarrow \infty$.
The flavour non-singlet axial-charges are renormalisation group invariants.

The isovector axial-charge is measured independently in neutron
$\beta$-decays
($g_A^{(3)} = 1.270 \pm 0.003$ \cite{PDG:2004})
and the octet axial charge is commonly taken 
to be the value extracted 
from hyperon $\beta$-decays assuming a 
2-parameter SU(3) fit
($g_A^{(8)} = 0.58 \pm 0.03$ \cite{fec}). 
The uncertainty quoted for $g_A^{(8)}$ has been a matter of some debate 
\cite{jaffe,ratcliffe}. 
There is 
considerable evidence that SU(3) symmetry may be badly broken 
and some have suggested that the error on $g_A^{(8)}$ should be 
as large as 25\%~\cite{jaffe}. 
Indeed, prompted by the work of Myhrer and Thomas~\cite{Myhrer:2007cf},
which showed that the 
effect of the one-gluon-exchange hyperfine interaction~\cite{Myhrer:1988ap}
and the pion cloud~\cite{Schreiber:1988uw}
of the nucleon was to reduce $g_A^{(0)}$ calculated in the cloudy bag model 
to near the experimental value, a re-evaluation of these effects on 
$g_A^{(3)}, \, g_A^{(8)}$ and $g_A^{(0)}$ including kaon loops led to 
the value $g_A^{(8)} = 0.46 \pm 0.05$~\cite{Bass:2009ed}.
Here the reduction 
from the SU(3) value came primarily from the pion cloud.

Assuming no twist-two subtraction constant, 
polarised deep inelastic scattering experiments have been interpreted 
in terms of a small value for the flavour-singlet axial-charge:
$
g_A^{(0)}|_{\rm pDIS, Q^2 \rightarrow \infty}
=
0.33 \pm 0.03 ({\rm stat.}) \pm 0.05 ({\rm syst.})
$
\cite{compassnlo}
if one uses the SU(3) value for $g_A^{(8)}$.
On the other hand, using the value $g_A^{(8)} = 0.46 \pm 0.05$ 
from SU(3) breaking,
the corresponding experimental value of 
$g_A^{(0)}|_{\rm pDIS}$ 
would increase to $g_A^{(0)}|_{\rm pDIS} = 0.36 \pm 0.03 \pm 0.05$. 
In the naive parton model $g_A^{(0)}|_{\rm pDIS}$ is interpreted 
as the fraction of the proton's spin which is carried by the intrinsic
spin of its quark and antiquark constituents.

Historically, the wish to understand the suppression of $g_A^{(0)}$ 
relative to $g_A^{(8)}$, 
led to considerable theoretical effort to understand
the flavour-singlet axial-charge in QCD.
QCD theoretical analysis leads 
to the formula~\cite{bassrmp,ar,et,ccm,bint}
\begin{equation}
g_A^{(0)}
=
\biggl(
\sum_q \Delta q - 3 \frac{\alpha_s}{2 \pi} \Delta g \biggr)_{\rm partons}
+ {\cal C}_{\infty}
.
\end{equation}
Here $\Delta g_{\rm partons}$ is the amount of spin carried
by polarised gluons in the polarised proton
($\alpha_s \Delta g \sim {\rm constant}$ as 
 $Q^2 \rightarrow \infty$~\cite{ar,et})
and
$\Delta q_{\rm partons}$ measures the spin carried by quarks
and
antiquarks
carrying ``soft'' transverse momentum $k_t^2 \sim P^2, m^2$
where
$P$ is a typical gluon virtuality
and
$m$ is the light quark mass.
The polarised gluon term is associated with events in polarised
deep inelastic scattering where the hard photon strikes a
quark or antiquark generated from photon-gluon fusion 
with
$k_t^2 \sim Q^2$~\cite{ccm,bint}.
It corresponds to the QCD axial anomaly in the flavour-singlet 
axial-vector current.
${\cal C}_{\infty}$ denotes a potential non-perturbative gluon
topological contribution~\cite{bassrmp}
associated 
with 
the possible subtraction 
constant in the dispersion relation for $g_1$ 
and 
possible spin contributions at Bjorken $x=0$, 
that is outside the range of polarised deep inelastic scattering experiments.
The measured singlet axial-charge is 
%full quantity minus the
%topological contribution:
$g_A^{(0)}|_{\rm pDIS} = g_A^{(0)} - C_{\infty}$.

In the parton model $\Delta q_{\rm partons}$ is associated with 
the forward matrix of 
the partially conserved axial-vector current $J_{+ 5}^{\rm con}$
evaluated in the light-cone gauge $A_+=0$
and corresponds to the quark spin contribution extracted 
from experiments using the JET or AB factorisation schemes \cite{jet}.
For each flavour $q$, 
this term and the possible topological term ${\cal C}_{\infty}$
are renormalisation group invariants \cite{bassrmp}.
All of the renormalisation group scale dependence induced by the anomalous 
dimension, $\gamma (\alpha_s)$, is carried by the polarised glue term:
\begin{equation}
\biggl\{
\frac{\alpha_s}{2 \pi} \Delta g \biggr\}_{Q^2}
=
\biggl\{
\frac{\alpha_s}{2 \pi} \Delta g \biggr\}_{\infty}
- \ \frac{1}{f} \
\biggl\{
1/E(\alpha_s) -1 \biggr\} \ g_A^{(0)} \bigg|_{\rm inv}
\label{eqf124}
\end{equation}
where all quantities in this equation are understood
to be defined in the $f$-flavour theory.
Flavour non-singlet combinations of the $\Delta q$ 
are renormalisation group invariant
so that each flavour evolves at the same rate, 
including heavy-quark contributions $(q=c,b,t)$.
The growth in the gluon polarisation
$
\Delta g \sim 1 / \alpha_s
$
at large $Q^2$ is compensated by growth with opposite sign in the
gluon orbital angular momentum. 
The quark and gluon total angular momenta in the infinite scaling 
limit are given by \cite{Ji}
$J_q ( \infty ) = \frac{1}{2} \{3 f / (16 + 3f)\}$ 
and
$J_g ( \infty ) = \frac{1}{2} \{16  / (16 + 3f)\}$.
There is presently a vigorous programme to disentangle the different
contributions involving experiments in semi-inclusive polarised deep
inelastic scattering and polarised proton-proton 
collisions~\cite{mpla,Mallot:2006}.

Heavy-quark axial-charges have been studied in the context of elastic 
neutrino-proton scattering \cite{bcst,Kaplan} 
and 
heavy-quark contributions to $g_1$ at $Q^2$ values above the charm 
production threshold \cite{AL,Manohar,bt92,bbs}.
Charm production in polarised deep inelastic scattering is an important 
part of the COMPASS spin programme at CERN 
\cite{charmexpt}.

Following Eq.(2) we can write the charm-quark axial-charge
contribution as
\begin{equation}
\Delta c (Q^2)
= \Delta c_{\rm partons} - 
\bigg\{ \frac{\alpha_s}{2 \pi} \Delta g \biggr\}_{Q^2,f=4}
\end{equation}
where $\Delta c_{\rm partons}$ corresponds to the forward matrix
element of the plus component of the 
renormalisation group invariant charm-quark axial-current
with just mass terms in the divergence (minus the QCD axial anomaly),
viz.
$({\bar c} \gamma_{\mu} \gamma_5 c)_{\rm con}
=
 ({\bar c} \gamma_{\mu} \gamma_5 c) - k_{\mu}$ 
with $k_{\mu}$ the gluonic Chern-Simons current, 
and we neglect any topological contribution
\footnote{Any topological contribution will be associated with 
 some of %the canonical 
 $\Delta c_{\rm partons}$ being shifted 
 to Bjorken $x=0$.
 In general, topological contributions
 are suppressed by powers of $1/m_c^2$ for heavy-quark matrix 
 elements \cite{shifman}.}.
For scales $Q^2 \gg m_c^2$, 
$\Delta c_{\rm partons}$ corresponds to the polarised charm
contribution one would find in the JET or AB factorisation schemes.

The heavy-quark contributions to the non-singlet neutral current
axial-charge measured in elastic neutrino-proton scattering have
been calculated to NLO in Ref.~\cite{bcst}.
For charm quarks, the relevant electroweak doublet contribution 
(at LO) is
\begin{eqnarray}
( \Delta c - \Delta s )_{\rm inv}^{(f=4)} 
&=&
- \frac{6}{27 \pi} \alpha_s^{(3)}(m_c^2)  g_A^{(0)}|_{\rm inv}^{(f=3)}
 - \Delta s_{\rm inv}^{(f=3)} 
\nonumber \\
& & 
+ \ 
O(1/m_c^2).
\end{eqnarray}
For the LO contribution this is made up from
\begin{eqnarray}
\Delta s_{\rm inv}^{(f=4)} 
&=& 
\Delta s_{\rm inv}^{(f=3)} 
+ (\frac{6}{27 \pi} - \frac{6}{25 \pi}) \alpha_s^{(3)} (m_c^2) 
g_A^{(0)}|_{\rm inv}^{(f=3)}
\nonumber \\
& & 
+ \ O(1/m_c^2) ,
\nonumber \\
\Delta c_{\rm inv}^{(f=4)} 
&=& 
- \frac{6}{25 \pi} \alpha_s^{(3)}(m_c^2) 
\
g_A^{(0)}|_{\rm inv}^{(f=3)} 
+ O(1/m_c^2) .
\label{eq:Deltac} \end{eqnarray}
Here $\Delta c_{\rm inv} = \Delta c (Q^2)$
evaluated in the limit $Q^2 \rightarrow \infty$,
where the charm-quark axial-charge contribution is
$
2 M s_{\mu} \Delta c =
\langle p,s | {\bar c} \gamma_{\mu} \gamma_5 c | p,s \rangle 
$.
The change in $\Delta s_{\rm inv}$ 
between the 4 and 3 flavour theories in Eq.(6) 
comes from the different
number of flavours in $E(\alpha_s)$ for the 4 and 3 flavour theories.

Eq.(6) contains vital information about
$\biggl\{ \frac{\alpha_s}{2 \pi} \Delta g \biggr\}_{\infty}$
in Eq.(4)
if we know the RG invariant quantity $\Delta c_{\rm partons}$.
Indeed, if the latter were zero and if we ignore the NLO evolution
associated with the two-loop anomalous dimension 
$\gamma (\alpha_s)$, 
then Eq.~(\ref{eq:Deltac}) would imply (at LO):
\begin{equation}
\Delta g^{(f=4)}(Q^2) 
= \frac{12}{25} \frac{\alpha_s^{(f=3)}(m_c^2)}{\alpha_s^{(f=4)}(Q^2)} 
g^{(0)}_A |^{(f=3)}_{\rm inv} \, ,
\end{equation}
or $\Delta g \sim 0.23$ when $\alpha_s(Q^2) \sim 0.3$. 
The following discussion 
is aimed at assessing the possible size of $\Delta c_{\rm partons}$ 
plus the NLO evolution associated with $\gamma (\alpha_s)$, and hence 
the error on this value.

Canonical (anomaly free)
heavy-quark contributions to the proton wavefunction are,
in general,
suppressed by powers of $1/m_c^2$, 
so we expect
$\Delta c_{\rm partons} \sim O(1/m_c^2)$.
The RG invariant quantity $\Delta c_{\rm partons}$ 
takes the same value at all momentum scales.
%
%is scale invariant. 
We may evaluate it 
using quark-hadron duality in a
hadronic basis with meson cloud methods \cite{wally}.
Experimental studies of the strange quark content of the nucleon 
over the last decade have given us considerable confidence that both 
the matrix elements of the vector and scalar charm quark currents 
(which are anomaly free) in the 
proton are quite  small~\cite{Young:2009zb,Young:2006jc}.
This gives us confidence in estimating the polarised charm contribution
through its suppression relative to the corresponding 
polarised strangeness 
-0.01 \cite{Bass:2009ed}
by the factor $\sim (m_\Lambda - m_N +m_K)^2/4m_c^2 < 0.1$ 
so that $|\Delta c_{\rm partons}| < 0.001$.
QCD 4-flavour evolution and Eq.(6)
then enables an estimate of 
$\Delta g$ at scales relevant to experiments at COMPASS and RHIC.

In perturbative QCD
the LO contribution to heavy-quark production through polarised 
photon-gluon fusion yields
\begin{equation}
\int_0^1 dx g_1^{(\gamma^* g \rightarrow h {\bar h})} \sim 0, 
\ \ \ Q^2 \gg m_h^2
\end{equation}
where $m_h$ is the heavy-quark mass ($h= c,b,t$).
The anomalous $- {\alpha_s / 2 \pi}$ term is cancelled against 
the canonical term when $m_h^2 \gg P^2$ 
(the typical gluon virtuality in the proton) \cite{ccm}.
If the gluon polarisation were large 
so that $-\frac{\alpha_s}{2 \pi} \Delta g$ 
made a large contribution to the suppression of $g_A^{(0)}$,
at this order
one would find also a compensating large canonical polarised 
charm contribution in the proton.
To understand this more deeply, we note that the result in Eq.(8) 
follows from the complete expression \cite{bint}
\begin{eqnarray}
& &
\int_0^1 dx \ g_1^{(\gamma^{*} g)}
= 
\nonumber 
\\
& &
- \frac{\alpha_s}{2 \pi}
\left[1 + \frac{2m^2}{P^2}
\frac{1}{\sqrt{ 1 + \frac{4 m^2}{P^2} } }
\ln \left(
\frac{\sqrt{1 + \frac{4 m^2}{P^2}} -1 }
{\sqrt{1 + \frac{4 m^2}{P^2}} +1 } \right) \right] .
\nonumber \\
\label{eqf122}
\end{eqnarray}
Here $m$ is the mass of the struck quark 
and $P^2$ is the gluon virtuality. 
We next focus on charm production.
The first term in Eq.(9) is the QCD anomaly and 
the second, mass-dependent, canonical term gives 
$\Delta c_{\rm partons}^{\rm (gluon)}$ 
for a gluon ``target" with virtuality $P^2$.
Evaluating Eq.(9) for $m_c^2 \gg P^2$ 
gives the leading term
$\int_0^1 dx \ g_1^{(\gamma^{*} g)} 
\sim 
- \frac{\alpha_s}{2 \pi} \frac{5}{8} \frac{P^2}{m_c^2}
$, -- hence the result in Eq.(8).

It is interesting to understand Eqs.(8-9) in terms of 
deriving the QCD axial-anomaly via Pauli-Villars regularisation 
(instead of the usual dimensional regularisation derivation 
 used in \cite{ccm}).
The anomaly corresponds to the heavy Pauli-Villars ``quark'',
which will cancel against the heavy charm-quark
for a charm-quark mass much bigger than gluon virtualities in 
the problem (there are no other mass terms to set the scale).
When the axial-vector amplitude is evaluated at two-loop level
there will be gluon loop momenta between $m_c$ and the ultraviolet
cut-off scale
generating a small scale dependence so that the cancellation 
between canonical heavy-quark and anomalous polarised glue terms 
is not exact in full QCD. 
This scale dependence corresponds to the two-loop anomalous
dimension $\gamma (\alpha_s)$ in $E(\alpha_s)$.
The result in Eq.(8) was previously discussed in 
Refs.\cite{jaffe,Manohar} 
in the context of the phenomenology that would 
follow if there were large gluon polarisation in the proton.
Non-perturbative evaluation of $\Delta c_{\rm partons}$
allows us to constrain the 
size of $\Delta g$ 
given what we know about charm and strangeness in the nucleon's
wavefunction.

There is a further issue that the derivation of Eq.(6) 
involves matching conditions where the spin contributions 
are continuous between the 3 and 4 
flavour theories at the threshold scale $m_c$ modulo $O(1/m_c^2)$
corrections, which determine a theoretical error for the method.
Using QCD evolution with the renormalisation group factor 
$E(\alpha_s)$, the results in Eq.(6) are equivalent 
to the leading twist term
$\Delta c (m_c^2)$
vanishing at the threshold scale $m_c$ modulo $O(1/m_c^2)$
corrections, 
viz.
$\Delta c (m_c^2) = O(1/m_c^2)$ \cite{jaffe}.
The leading $O(1/m_c^2)$ term is estimated using effective field
theory in Refs.~\cite{jaffe,Kaplan,Manohar}. 
For polarised photon-gluon fusion, this is the 
$-\frac{\alpha_s}{2 \pi} \frac{5}{8} \frac{P^2}{m_c^2}$
leading term in the heavy-quark limit of Eq.(9).
The heavy charm-quark is integrated out at threshold
to give the matrix element of a gauge invariant gluon
operator with dimension 5 and the same quantum numbers 
as the axial-vector current, 
viz. 
$
\Delta c (m_c^2) 
\sim O \biggl(
\frac{\alpha_s (m_c^2)}{4 \pi} \frac{M^2}{m_c^2}
 \biggr)
$
\cite{Manohar}
%\end{equation}
or
$\Delta c (m_c^2) 
 \sim O( \alpha_s (m_c^2) {\Lambda_{\rm qcd}^2 / m_c^2} ) 
 \sim 0.017$
\cite{jaffe}
\footnote{
These $O(1/m_c^2)$ terms associated with the full $\Delta c$
 are manifest in polarised photon-gluon fusion through 
 the heavy-quark limit of Eq.(9),
 $\int_0^1 dx \ g_1^{(\gamma^{*} g)} 
 \sim 
 - \frac{\alpha_s}{2 \pi} \frac{5}{8} \frac{P^2}{m_c^2}$,
 and are to be distinguished from the model evaluation 
 of $\Delta c_{\rm partons}$.}.
Taking this as an estimate of the theoretical error gives
$\Delta c (m_c^2) = 0 \pm 0.017$.

We next combine this number for $\Delta c (m_c^2)$
with our
estimate of the canonical charm contribution 
$|\Delta c_{\rm partons}| < 0.001$
in quadrature 
to obtain a bound including 
theoretical error on the size of 
the polarised gluon contribution:
$| - \frac{\alpha_s}{2 \pi} \Delta g (m_c^2) |
\lesssim
0.017$
or 
\begin{equation}
| \Delta g (m_c^2) | \lesssim 0.3
\end{equation}
with $\alpha_s (m_c^2) = 0.4$.
Values at other values of $Q^2$ are readily obtained with Eq.(3)
or $\alpha_s \Delta g \sim {\tt constant}$ for large values of $Q^2$.

It is interesting to extend this analysis to full 6-flavour QCD.
The values of 
$\Delta c_{\rm inv}^{(f=6)}$, 
$\Delta b_{\rm inv}^{(f=6)}$ and $\Delta t_{\rm inv}^{(f=6)}$
were derived in Ref. \cite{bcst} to NLO in the heavy-quark expansion.
Taking just the leading-order contribution plus the heavy-quark power 
correction according to the recipe \cite{jaffe,Manohar} described above
gives
$\Delta c^{(f=5)}(m_b^2) = -0.006 \pm 0.017$ 
and $\Delta c^{(f=6)}(m_t^2) = -0.009 \pm 0.017$ for polarised charm.
For the bottom and top quarks one obtains
$\Delta b^{(f=5)}(m_b^2) = 0 \pm 0.001 \pm 0.017$, 
$\Delta b^{(f=6)}(m_t^2) = -0.003 \pm 0.001 \pm 0.017$
and 
$\Delta t(m_t^2) = 0 \pm 2 \times 10^{-7} \pm 0.017$.
Here the first error comes from the $O(1/m_h^2)$ mass correction 
for the heaviest quark of $c, b, t$.
The second error comes from the other heavy-quarks with lesser mass
as we evaluate these heavy-quark contributions in terms of the measured
value of the light-quark quantity $g_A^{(0)}|_{\rm inv}^{(f=3)}$.
These numbers
overlap with a 
zero value for $\frac{\alpha_s}{2 \pi} \Delta g$ in the relevant 
$f-$flavour theories.
The QCD scale dependence of $\frac{\alpha_s}{2 \pi} \Delta g$ 
starts with NLO evolution induced 
by Kodaira's two-loop anomalous dimension $\gamma (\alpha_s)$. 
The combination $\frac{\alpha_s}{2 \pi} \Delta g$ is scale 
invariant at LO. This means that if we work just to LO and
$\Delta g$ vanishes at one scale, it will vanish at all scales 
(in LO approximation).
The LO QCD evolution equation for gluon orbital angular momentum 
in the proton \cite{Ji} then simplifies so that
$L_g ( \infty ) = \frac{1}{2} \{16  / (16 + 3f)\}$.
In practice, 
the two-loop anomalous dimension generates slow evolution of 
$\frac{\alpha_s}{2 \pi} \Delta g$.
Dividing the finite value of this combination at large scales 
by the small value of $\alpha_s$ gives a finite value for the 
gluon polarisation $\Delta g$, which can readily be the same
order of magnitude as the 
gluon total angular momentum 
(or larger with cancellation against a correspondingly larger gluon
 orbital angular momentum contribution).

It is interesting that the value of $\Delta g$
deduced from present experiments COMPASS at CERN and PHENIX and
STAR at RHIC typically give
$|\Delta g| < 0.4$ with $\alpha_s \sim 0.3$
corresponding to 
$|- 3 \frac{\alpha_s}{2 \pi} \Delta g| < 0.06$ \cite{mpla}.
This experimental value is extracted from direct measurements of
gluon polarisation at COMPASS in the region around 
$x_{\rm gluon} \sim 0.1$, NLO QCD motivated fits to inclusive 
$g_1$ data taken in the region $x > 0.006$, and RHIC Spin data in the region 
$0.02 < x_{\rm gluon} < 0.4$.
The theoretical bound, Eq.(10), is consistent with this experimental result.

\vspace{0.5cm}

{\it Acknowledgements.} 
The research of
SDB is supported by the Austrian Science Fund, FWF, through grant 
P20436, 
while
AWT is supported by the Australian Research Council through an 
Australian Laureate Fellowship and by the University of Adelaide.

\end{document}